\documentclass[longbibliography,twocolumn,reprint,a4paper]{revtex4-1}

\usepackage{amsmath}
\usepackage{amssymb}
\usepackage{graphicx,color}

\usepackage{lineno}
\begin{document}

\title{A MHD reverse flow in 90 degree bend}

\author{Alexander~V. Proskurin}
\affiliation{Altai State Technical University, 656038, Russian Federation, Barnaul, Lenin prospect,46}
\email{k210@list.ru}

\author{Anatoly~M. Sagalakov}
\affiliation{Altai State University}

\date{\today}

\begin{abstract}
A two-dimensional flow in a 90 degree bent channel is considered. A magnetic field is uniform and parallel to inlet branch of the channel. A spectral/hp element method was used for liquid motion calculations. Three types of steady flows were detected. It was found that magnetic forces can suppress a pressure gradient and throw out liquid from the inlet, a relatively large reverse flow appears.
\end{abstract}


\maketitle

\sloppy


Advanced fusion reactors use blankets cooled by liquid metals. The thermonuclear plasma is held by a magnetic field, which strongly influences to liquid metal flows in blankets. MHD flows in straight pipes are well studied in the laminar case, in transient regimes, in the case of developed turbulence. In real life devices, the fluid flows not only straight, but also changes direction. It seems interesting to identify several typical geometric configurations and study them in detail. Accumulated knowledge of these flows will be useful in the design of liquid metal facilities.

\begin{figure}[ht]
\begin{center}
\includegraphics{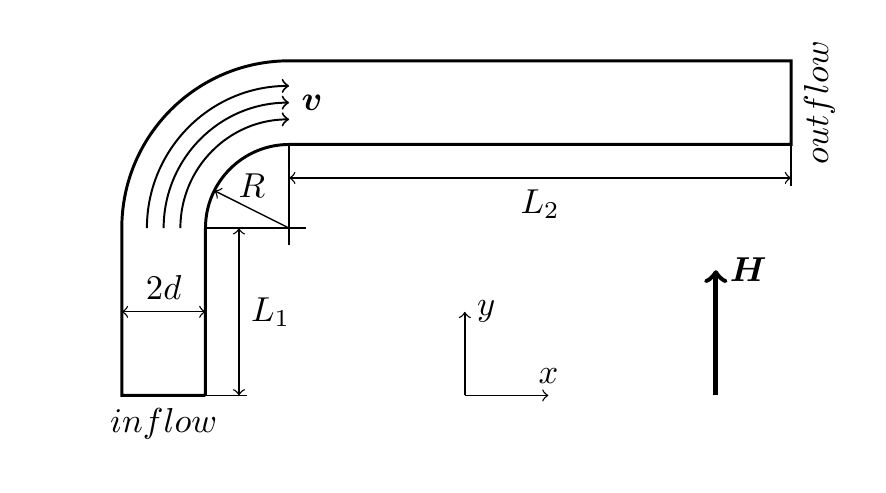} 
\end{center}
\caption{A 90 degree bent channel}\label{article21.LShapeGeometry}
\end{figure}

A common example of such flows is a flow in bent pipe. this flow in the case of a nonconducting fluid is well investigated. According to McNally\cite{spedding2004fluid}, the bent pipes can be classified: curved pipes, elbow pipes, U-shaped pipes. Numerous studies of curved pipes were started by Dean \cite{dean1927xvi}, in which work the analytical expressions for the flow were proposed. U-shaped pipes are also well explored, since this pipes are interesting as heat exchangers (see the bibliography in review \cite{spedding2004fluid}).
Laminar regimes in elbow pipes are studied worse\cite{van1989finite,pantokratoras2016steady}.

We consider a two-dimensional flow in a 90 degree bent channel, a sketch is shown at Figure \ref{article21.LShapeGeometry}. A fluid is assumed to be viscous and electrically conductive. A magnetic field is directed vertically. The fluid flows into the channel through the "inflow" section and leaves channel through the "outflow" section of the boundary. It is possible to define a aspect ratio parameter that determines geometry of the channel:
\begin{equation}
\delta = \frac{d}{R}.
\end{equation}

Consider magnetohydrodynamics equations in the form
\begin{equation}
\label{article21.NS_Syst_MagForce}
\begin{aligned}
\frac{\partial \boldsymbol{v}}{\partial{t}}+\left( \boldsymbol{v} \nabla \right)\boldsymbol{v} &= -\frac{1}{\rho}\nabla p + \nu \Delta \boldsymbol{v} + \boldsymbol{F}(\boldsymbol{v},\boldsymbol{H}),\\
div \boldsymbol{v} &= 0,
\end{aligned}
\end{equation}
where $\boldsymbol{v}$ is the velocity, $p$ is the pressure, $\nu$ is the viscosity, $\rho$ is the density, $\boldsymbol{F}$ is the magnetic force, and $\boldsymbol{H}$ is the magnitude of the imposed magnetic field.

Ohm's law is:
\begin{equation}
\label{article21.j_eq}
\boldsymbol{j} = \sigma\left( -\nabla\varphi+\boldsymbol{v}\times\boldsymbol{H}  \right),
\end{equation}
where $\boldsymbol{j}$ is the density of electric current, $\varphi$ is the electric potential, and $\sigma$ is the conductivity. Using the law of conservation of electric charge ($div \boldsymbol{j} = 0$), it is possible to derive the equation for electric potential as:
\begin{equation}
\label{article21.ElPot_eq}
\Delta \varphi = \nabla(\boldsymbol{v}\times\boldsymbol{H}).
\end{equation}

The system (\ref{article21.NS_Syst_MagForce}) can be written in the form: 
\begin{equation}
\label{article21.WeakMHD_Syst}
\begin{aligned}
\frac{\partial \boldsymbol{v}}{\partial{t}}+\left( \boldsymbol{v} \nabla \right)\boldsymbol{v} &= -\nabla p + \frac{1}{Re} \Delta \boldsymbol{v} +\\
&+ St \left( -\nabla\varphi+\boldsymbol{v}\times\boldsymbol{H}  \right)\times\boldsymbol{H} ,\\
&div \boldsymbol{v} = 0,\\
&\Delta \varphi = \nabla(\boldsymbol{v}\times\boldsymbol{H}),
\end{aligned}
\end{equation}
where $Re = \frac{dV_0}{\nu}$ is the Reynolds number, $St=\frac{\sigma H_0^2 d}{\rho V_0}=\frac{Ha^2}{Re}$ is the magnetic interaction parameter (Stuart number), $Ha$ is Hartmann number, and $d$(the half width of the channel), $V_0$, and $H_0$ represent the scales of length, velocity and magnitude of the imposed magnetic field, respectively. This system is widely used in theoretical investigations and accurately approximates many cases of liquid metal flows 
\cite{krasnov2011comparative}.

The boundary condition for velocity at walls have the form:
\begin{equation}
\label{article21.BounCond_V}
\boldsymbol{v} = 0.
\end{equation}
It is supposed that the flow is two-dimensional, all functions are independent from $z$-coordinate and $v_z=0$. This proposition leads to the equation $\Delta \varphi = 0$,  that is mean that without external electric potentials $\varphi = 0$ everywhere and boundary conditions for $\varphi$ is not required.

\begin{table}[h]
\caption{Velocity convergence at points A, B and C}\label{article21.convergence_table}
\begin{center}
\tiny
\begin{tabular}{|r||c|c|c||c|c|c|c|}
\hline
 &\multicolumn{3}{|c|}{rectilinear grid}&\multicolumn{3}{|c|}{truangular grid}\\
\hline
$p$ & $V_x$ at A & $V_y$ at B & $V_x$ at C & $V_x$ at A & $V_y$ at B & $V_x$ at C \\
\hline
3 &-0.00614057 &-0.03366661 &0.01152448 &-0.00252486 &-0.04073437 &-0.0065677 \\
5 &-0.00046651 &-0.03109454 &0.00989152 &-0.00104130 &-0.03170791 &0.01107090 \\
7 &-0.00078967 &-0.03163357 &0.01168906 &-0.00101746 &-0.03169345 &0.01162540 \\
9 &-0.00104049 &-0.03167006 &0.01164027 &-0.00103455 &-0.03173050 &0.01163066 \\
11&-0.00102927 &-0.03167162 &0.01159739 &-0.00103392 &-0.03170722 &0.01161798 \\
13&-0.00103334 &-0.03166994 &0.01159802 &-0.00103287 &-0.03168602 &0.01160666 \\
15&-0.00103240 &-0.03166936 &0.01159777 &-0.00103264 &-0.03167486 &0.01160077 \\
17&-0.00103254 &-0.03166915 &0.01159760 &-0.00103257 &-0.03167044 &0.01159842 \\
\hline
\end{tabular}
\end{center}
\end{table}

\begin{figure}[h]
\begin{center}
\includegraphics[width=0.5\textwidth]{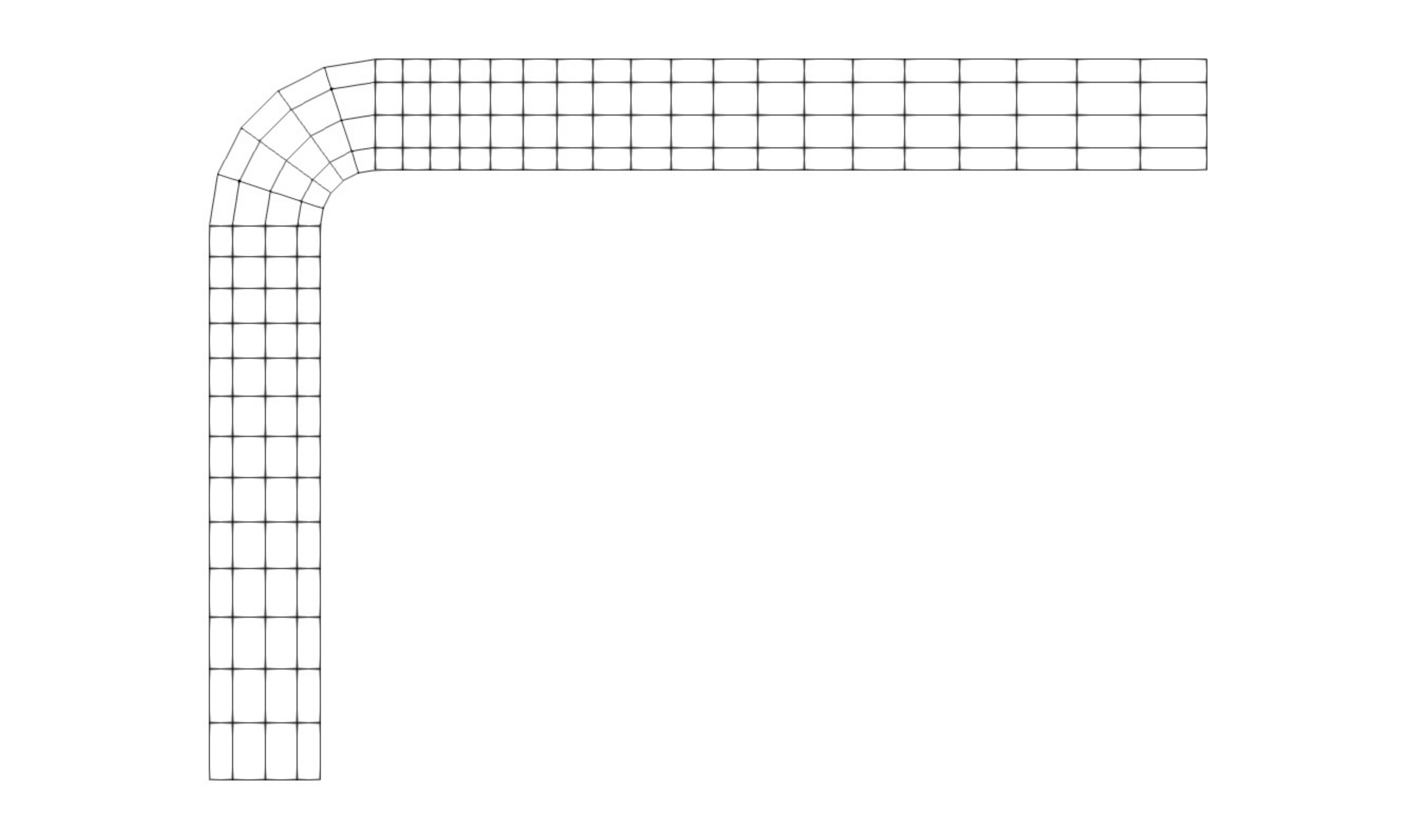}
\end{center}
\caption{The rectilinear grid}\label{article21.grid_rect}
\end{figure}

\begin{figure}[h]
\begin{center}
\includegraphics[width=0.5\textwidth]{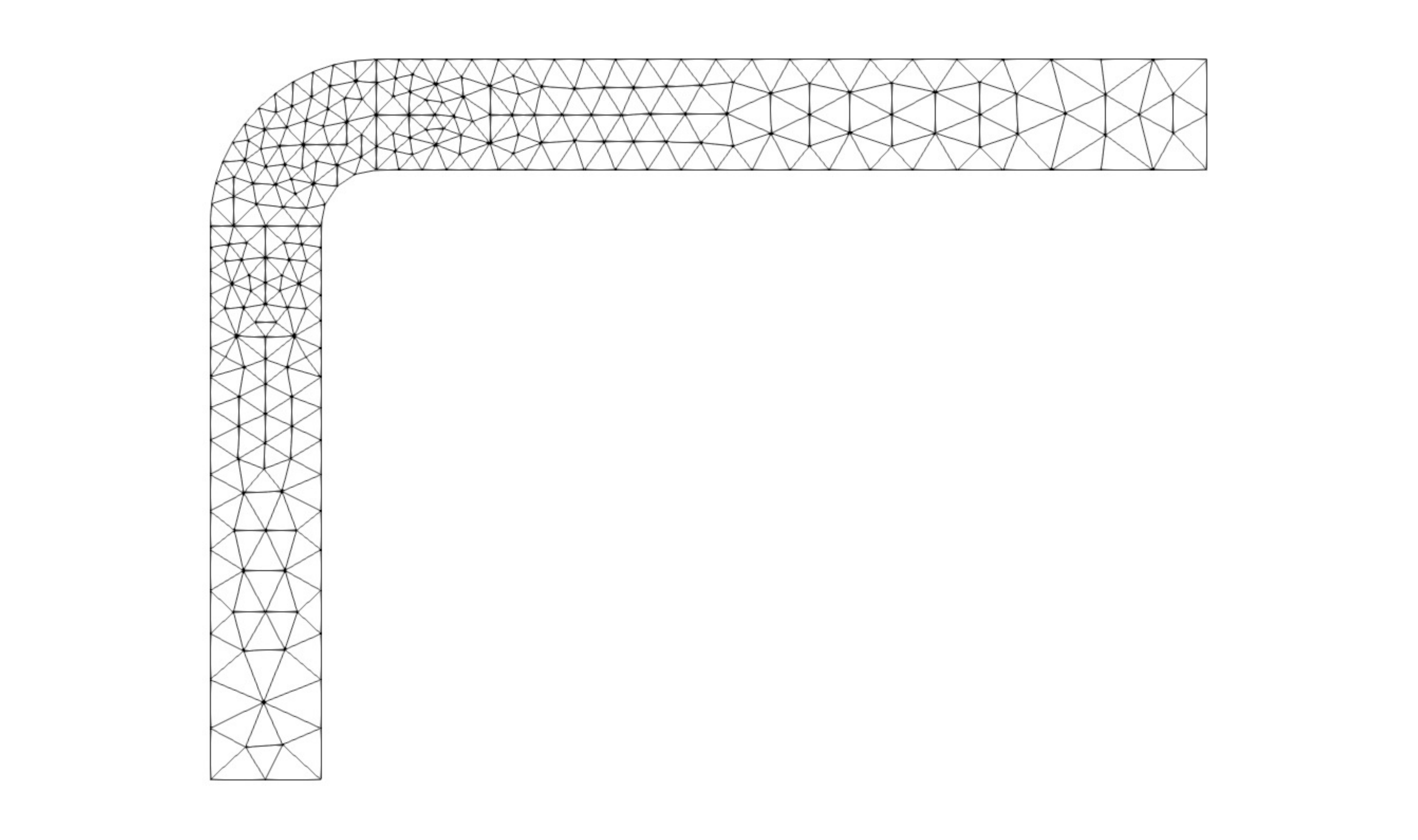}
\end{center}
\caption{The triangular grid}\label{article21.grid_tri}
\end{figure}


We will investigate steady flows. To do this, we set a constant pressure gradient between the channel's inflow and outflow and zero initial conditions. The MHD solver has been developed on the basise of an open source spectral/hp element framework \texttt{Nektar++} \cite{cantwell:2015, karniadakis:2013}.
The incompressible Navier-Stokes solver (\texttt{IncNavierStokesSolver}) from this framework has been taken as the source for the MHD solver. Reliability and convergence of numeric method is discussed in \cite{proskurin2017spectral}.

\begin{figure}
\begin{center}
\includegraphics[width=0.3\textwidth]{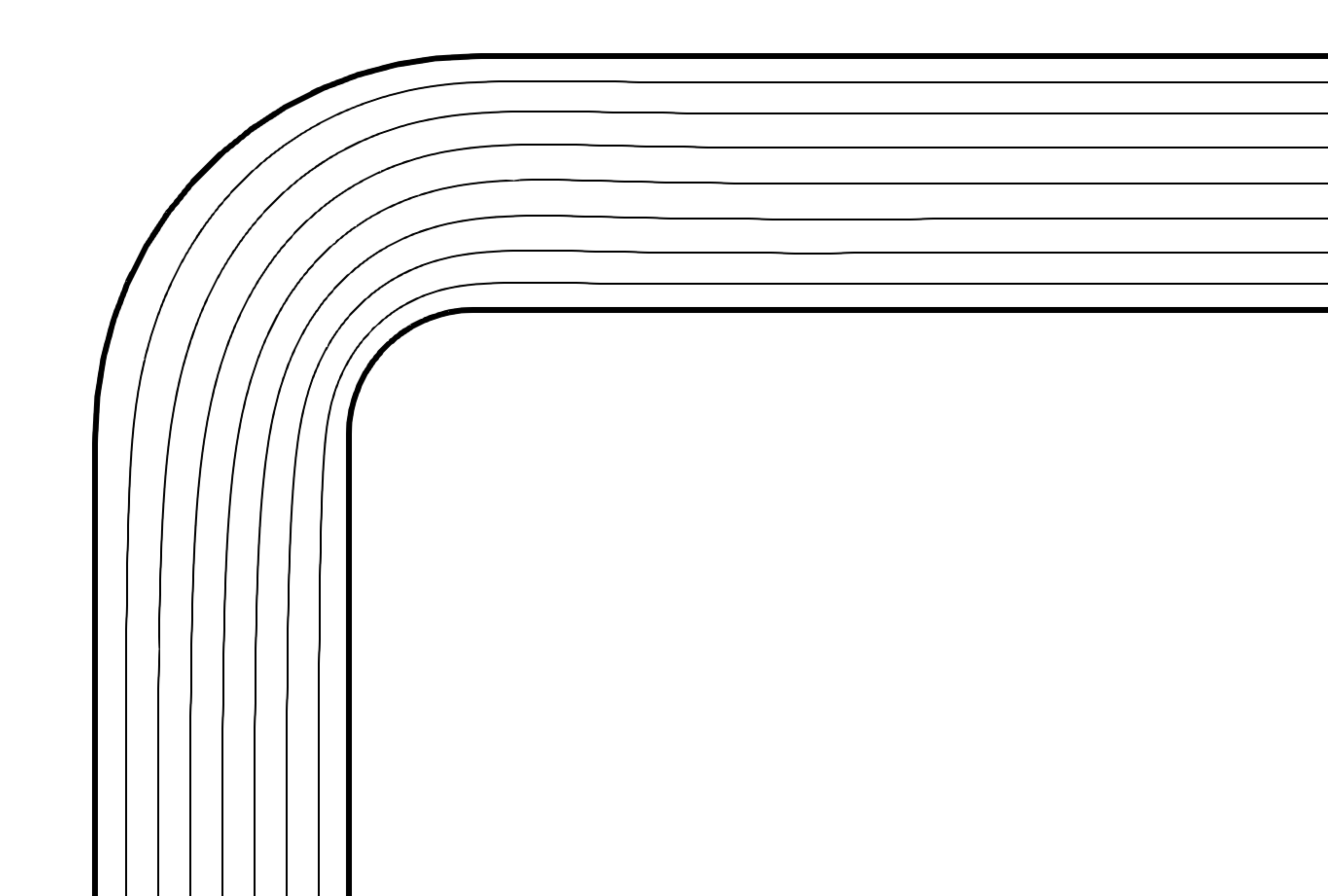}
\end{center}
\caption{Streamlines of the Type0 flow, $Re=4.85$, $St=2.06$, $\delta=1$}\label{article21.flow_type0}
\end{figure}

\begin{figure}
\begin{center}
\includegraphics[width=0.3\textwidth]{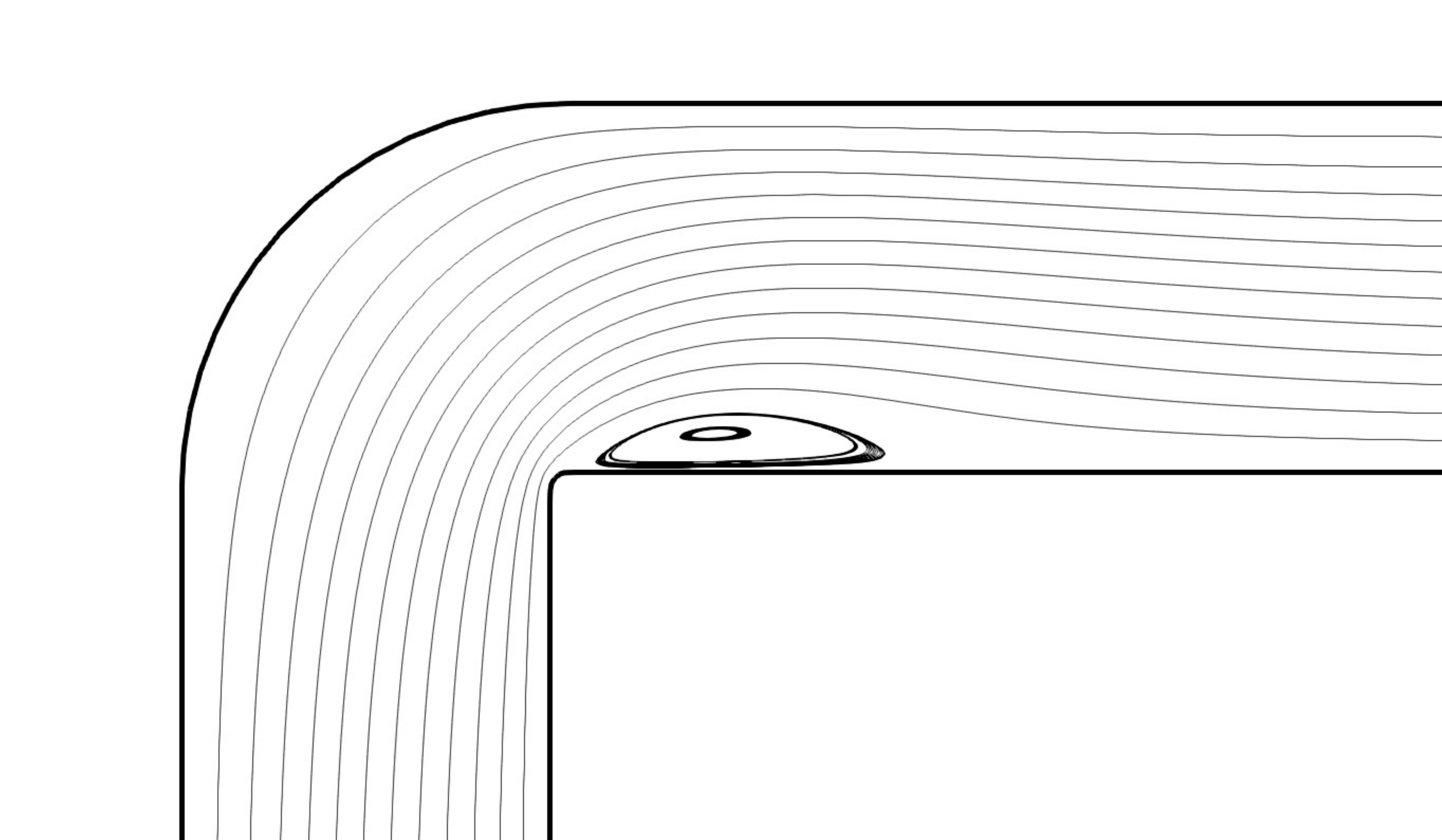}
\end{center}
\caption{Streamlines of the Type1 flow, $Re=50$, $St=0.001$, $\delta=0.1$}\label{article21.flow_type1}
\end{figure}

\begin{figure}
\begin{center}
\includegraphics[width=0.3\textwidth]{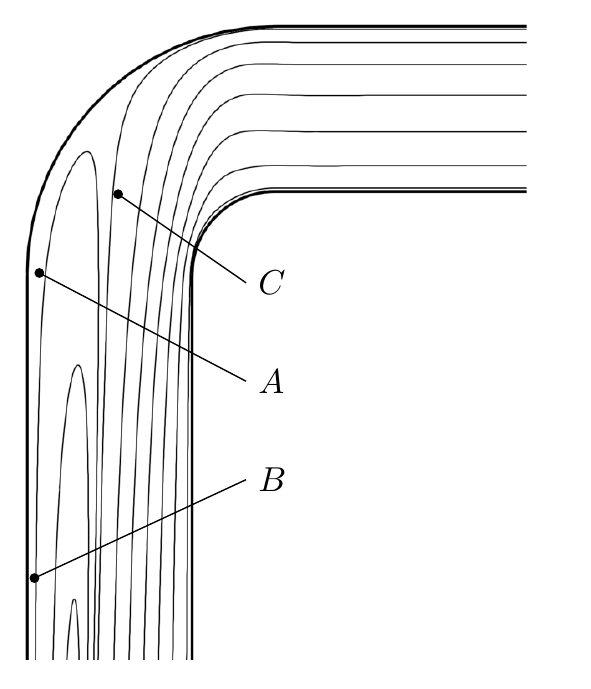}
\end{center}
\caption{Streamlines of the Type2 flow, $Re=103.16$, $St=58.16$, $\delta=1$}\label{article21.flow_type2b}
\end{figure}

Figures \ref{article21.grid_rect} and \ref{article21.grid_tri} show rectangular and a triangular grids that have 251 and 561 elements, respectively. For mass production calculations, the rectangular grid was used. For convergence control we used the triangular grid. The velocity values was controlled at several points A, B, C, whose location is shown at Figure \ref{article21.flow_type2b}. Table \ref{article21.convergence_table} lists this values for different orders of polynomial approximation $p$. The numeric method has convergence at least 4 digits using the rectangular grid. In the case of the triangular grid converges is less rapid. The table allows us to conclude that the calculation error is less than $0.01\%$.

\begin{figure}
\begin{center}
\includegraphics[width=0.3\textwidth]{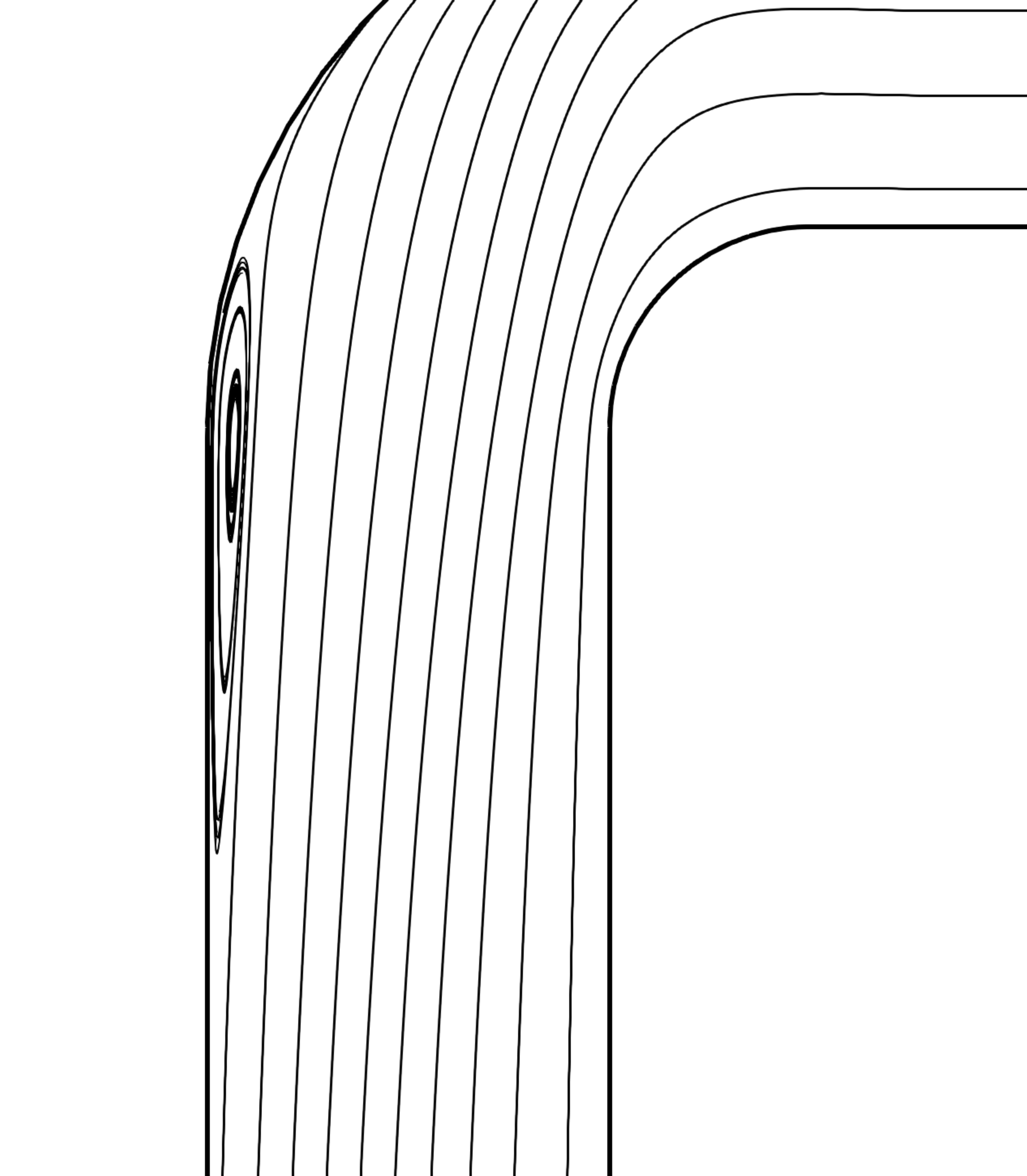}
\end{center}
\caption{Streamlines of the Type2 flow, $Re=25.63$, $St=58.54$, $\delta=1$}\label{article21.flow_type2a}
\end{figure}

\begin{figure}
\begin{center}
\includegraphics[width=0.35\textwidth]{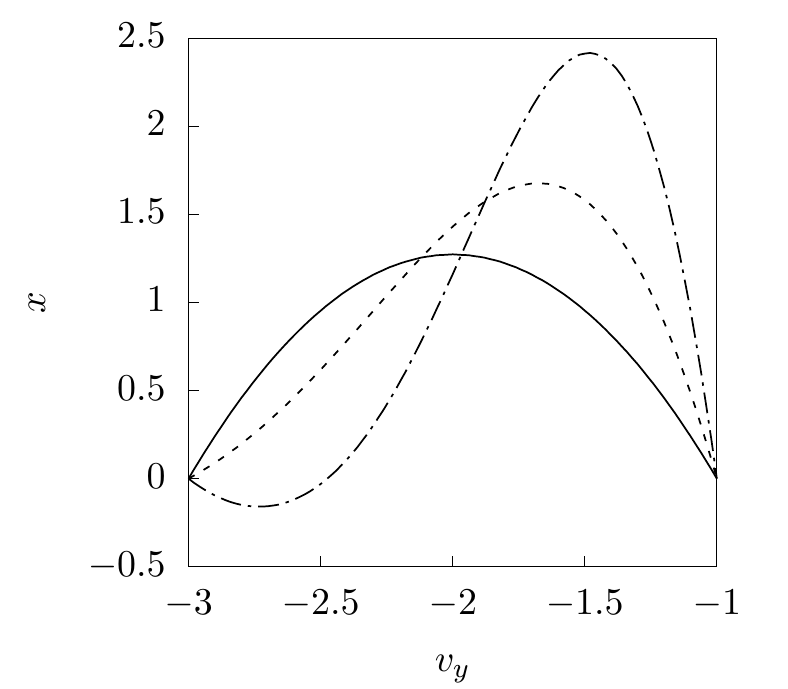}
\end{center}
\caption{Type2 velocity plots near the inlet,  $Re=20$, $St=59$(solid line), $Re=64$, $St=62$(dashed line), $Re=103$, $St=58$(pointed-dashed line), $\delta=1$}\label{article21.profile_type2_1}
\end{figure}

\begin{figure}
\begin{center}
\includegraphics[width=0.35\textwidth]{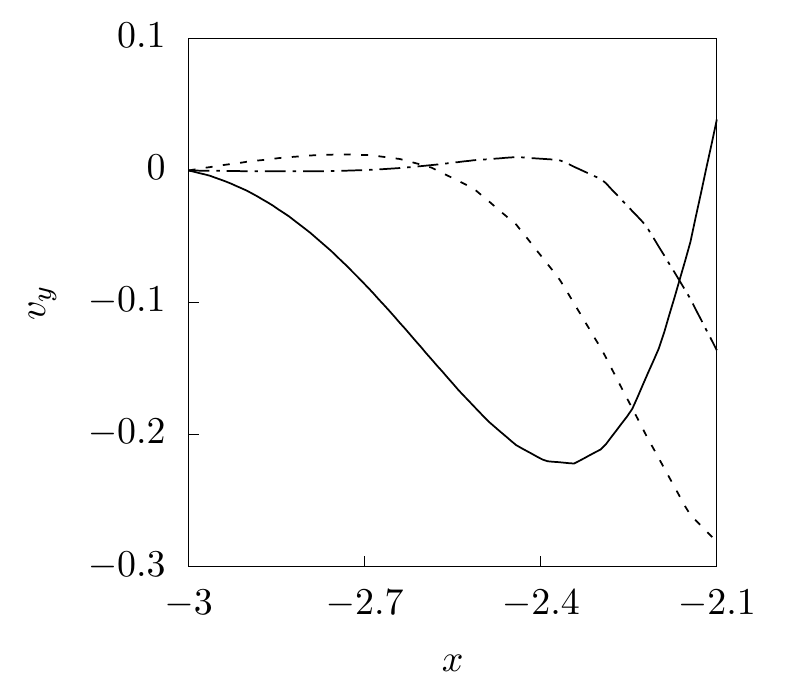}
\end{center}
\caption{Enlarged view of Type2 velocity plots near the inlet, $Re=518$, $St=58$(solid line), $Re=2080$, $St=58$(dashed line), $Re=13023$, $St=58$(pointed-dashed line), $\delta=1$}\label{article21.profile_type2_2}
\end{figure}

\begin{figure}
\begin{center}
\includegraphics[width=0.35\textwidth]{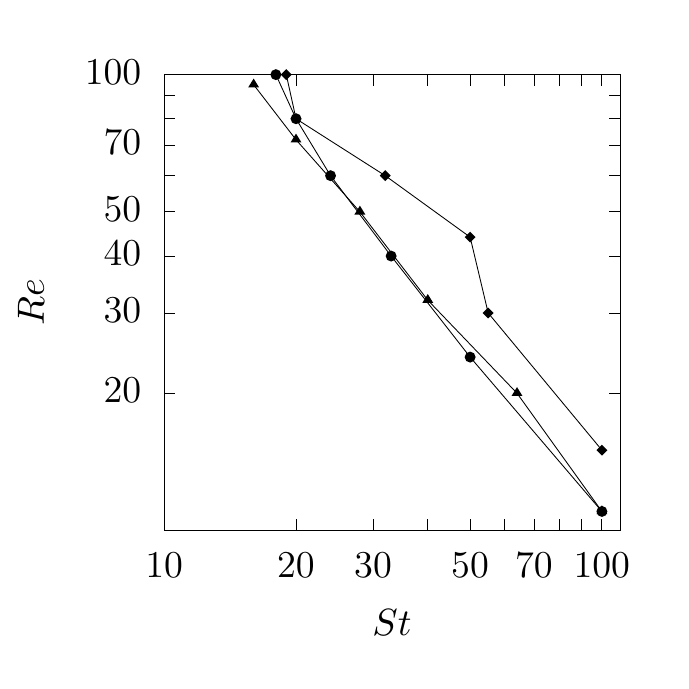}
\end{center}
\caption{Border lines Type0$\rightarrow$Type2 at $\delta=0.1$(circle), $\delta=1$(triangle), $\delta=5$(diamond)}\label{article21.diagram_T0_to_T2}
\end{figure}

\begin{figure}
\begin{center}
\includegraphics[width=0.35\textwidth]{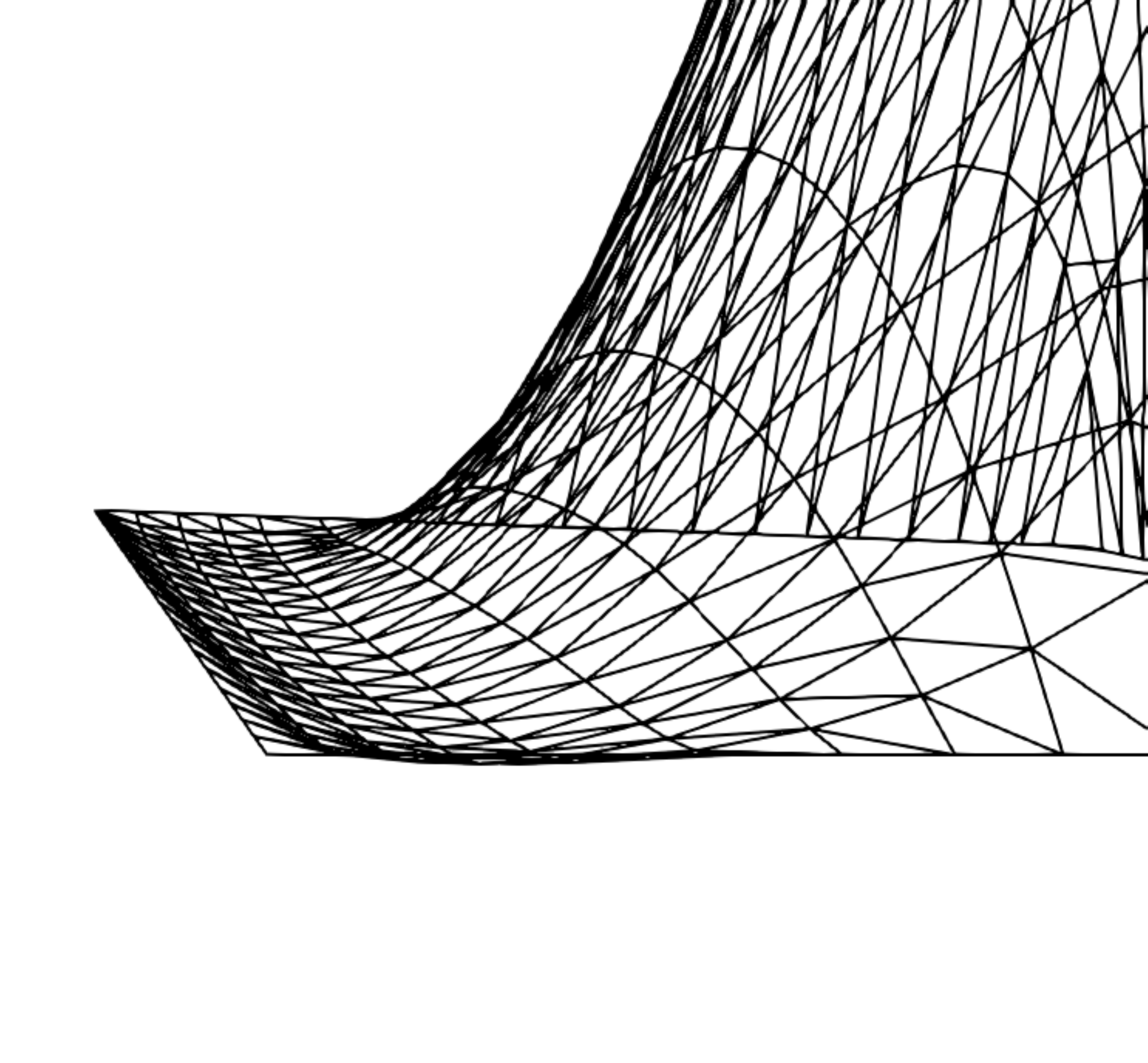}
\end{center}
\caption{Velocity plot near the inlet in three-dimensional case, $Re=103$, $St=58$}\label{article21.profile_3d}
\end{figure}


We investigated the flow in a wide range of Reynolds and Stuart numbers. It was found that the flow has three different steady patterns. The first pattern is a parallel flow marked as Type0 (Figure \ref{article21.flow_type0}). The second pattern (Type1) has a vortex in the outlet branch right after the elbow (Figure \ref{article21.flow_type1}). Type0 and Type1 patterns are well known and can be observed without magnetic field. If the viscosity is decreased under influence of sufficiently large magnetic forces, the third flow pattern arises (Type2, Figure \ref{article21.flow_type2a}). In case if the viscosity is decreased further, the vortex in the inlet branch can turn into a reverse flow (Figure \ref{article21.flow_type2b}). Plots of velocity in the inflow branch of channel are shown in Figures \ref{article21.profile_type2_1} and \ref{article21.profile_type2_2}. For these calculations, we changed the viscosity so that the Reynolds number varies from $20$ to $13023$. The electrical conductivity was set constant. A origin of reverse flow can be observed in Figure \ref{article21.profile_type2_1}. Figure \ref{article21.profile_type2_2} illustrates dislocation of this reverse jet to the center line of the inlet branch. Additionally, we changed both the conductivity and the viscosity and investigated transition from Type0 to Type1. A diagram shown in Figure \ref{article21.diagram_T0_to_T2} illustrates this transition for $\delta=0.1,\,1,\,5$. The reverse flow phenomenon is also observed in three-dimensional case. The authors carried out pilot calculations for a square cross section pipe. The velocity graph near the entrance is shown in Figure \ref{article21.profile_3d}.


Reverse flows frequently appear as vortices behind obstacles, such as a cylinder in homogeneous flow or a step in channel.
In the bent channel such vortex can be found after elbow (see Figure \ref{article21.flow_type1}). Also reverse flows can occur if there are forces acting opposite to a main flow current, an example is presented by Abdou \cite{SMOLENTSEV2008771}.
In the considered in article case, the appearance of reverse flow is unexpected, since it does not follow from obvious features of the magnetic field or the channel geometry. The authors suggest that this effect can have a strong influence on flows of liquid metals in various devices, such as blankets of thermonuclear reactors.

\bibliography{reference21}

\end{document}